\documentclass[11pt,a4paper]{PoS}

\newcommand{\Fig}[1]{Fig.~\ref{#1}}
\newcommand{\Ref}[1]{Ref.~\cite{#1}}
\newcommand{\Eq}[1]{Eq.~(\ref{#1})}

\newcommand{\mean}[1]{\ensuremath{\langle#1\rangle}}
\newcommand{\order}[1]{\ensuremath{\mathcal{O}(#1)}}
\newcommand{\SN}[2]{\ensuremath{#1\times10^{#2}}}

\newcommand{\Eg}{\ensuremath{E_\gamma^*}}
\newcommand{\pe}{\ensuremath{p_e^*}}
\newcommand{\qg}{\ensuremath{\theta_\gamma^*}}

\title{Recent KLOE results on radiative kaon decays}

\ShortTitle{Recent KLOE results on radiative kaon decays}

\author{KLOE Collaboration\thanks{
F.~Ambrosino, A.~Antonelli, M.~Antonelli, F.~Archilli, P.~Beltrame,
G.~Bencivenni, C.~Bini, C.~Bloise, S.~Bocchetta, F.~Bossi, P.~Branchini,
G.~Capon, D.~Capriotti, T.~Capussela, F.~Ceradini, P.~Ciambrone, E.~De Lucia,
A.~De Santis, P.~De Simone, G.~De Zorzi, A.~Denig, A.~Di Domenico,
C.~Di Donato, B.~Di Micco, M.~Dreucci, G.~Felici, S.~Fiore, P.~Franzini,
C.~Gatti, P.~Gauzzi, S.~Giovannella, E.~Graziani, M.~Jacewicz, V.~Kulikov,
G.~Lanfranchi, J.~Lee-Franzini, M.~Martini, P.~Massarotti, S.~Meola, 
S.~Miscetti, M.~Moulson, S.~M\"uller, F.~Murtas, M.~Napolitano, F.~Nguyen,
M.~Palutan, A.~Passeri, V.~Patera, P.~Santangelo, B.~Sciascia, A.~Sibidanov,
T.~Spadaro, M.~Testa, L.~Tortora, P.~Valente, G.~Venanzoni, and R.~Versaci.}\\
\speaker{presented by Matthew Moulson}\\
         INFN Frascati, Italy\\
         E-mail: \email{moulson@lnf.infn.it}
}

\abstract{
While measuring the ratio  
$R_K = \Gamma(K^\pm_{e2(\gamma)})/\Gamma(K^\pm_{\mu2(\gamma)})$,
the KLOE Collaboration has studied the radiative process $K_{e2\gamma}$.
The ratio of the width for the $K_{e2\gamma}$ decay with a positively
polarized photon from structure-dependent radiation to the inclusive
$K_{\mu2(\gamma)}$ width is found to be \SN{1.484(68)}{-5}. The
observed radiation spectrum agrees with predictions from chiral 
perturbation theory and is in contrast with predictions based on the 
light front quark model. This result reduces the contribution to 
systematic uncertainties on measurements of $R_K$. In a separate
study, KLOE has measured the ratio of the radiative $K_{e3\gamma}$ 
decay width to the inclusive $K_{e3(\gamma)}$ width to be \SN{924(28)}{-5}.
The distribution in energy and angle of the radiative photon has 
been analyzed in an attempt to isolate the signature from interference
of the inner-bremsstrahlung and structure-dependent amplitudes.}

\FullConference{2009 KAON International Conference KAON09,\\
		 June 09 - 12 2009\\
		 Tsukuba, Japan}

\begin{document}

\section{$K^\pm\to e^\pm\nu(\bar\nu)\gamma$ $(K_{e2\gamma})$}

The $K_{e2}$ decay is strongly helicity suppressed. Its rate is therefore
a sensitive probe for minute contributions from physics beyond the 
Standard Model (SM). This is particularly true of the
ratio $R_K=\Gamma(K_{e2(\gamma)})/\Gamma(K_{\mu2(\gamma)})$,
which can be calculated in the SM without uncertainties from
strong-interaction dynamics.
A recent calculation, which includes \order{e^2p^4}
corrections in chiral perturbation theory (ChPT), gives 
$R_K^{\rm SM} = \SN{2.477(1)}{-5}$ \cite{CR07:Kp2}.
Deviations of $R_K$ of up to a few percent are possible in minimal
supersymmetric extensions of the SM with non-vanishing $e$-$\tau$
scalar-lepton mixing \cite{MPP06:Ke2}. 

Additional photons in the $K_{e2}$ final state can be produced via
internal bremsstrahlung (IB) or structure-dependent (direct) emission (SD);
the spectrum of the latter is sensitive to the structure of the kaon.  
Interference between the IB and SD processes is negligible \cite{B+95:l2Hand2}.
By definition, $R_K$ is IB inclusive and SD exclusive.
However, since events with IB photons cannot be distinguished from
those with SD photons, to compare data with the SM prediction at the
percent level or better, sufficiently precise knowledge of the 
SD component is required.

Because of the helicity suppression of the IB channel, the contribution 
from the SD channel to the total width is approximately equal to that from IB 
\cite{B+95:l2Hand2}. Previous measurements give the SD rate with a
relative uncertainty of 15\% \cite{H+79:Ke2}. The KLOE Collaboration
has recently submitted for publication a measurement of $R_K$ to within 
1.3\% \cite{KLOE+09:Ke2,Sci09:Kaon}, obtaining $R_K = \SN{2.493(31)}{-5}$.
The study of the SD radiation in $K_{e2\gamma}$ decay presented here was 
conducted in order to reduce the contribution to the systematic error
on the KLOE measurement of $R_K$ from the uncertainty in the SD rate,
from 0.5\% to 0.2\%. The NA62 Collaboration has recently announced a 
preliminary measurement of $R_K$ with a 0.6\% error \cite{Gou09:Kaon}.
The uncertainty in the SD rate is the second-largest contribution to
the overall systematic uncertainty on this result, so that the present
study may help to reduce the uncertainty on the NA62 value of 
$R_K$ as well.

\begin{figure}
\centerline{
\includegraphics[width=0.33\textwidth]{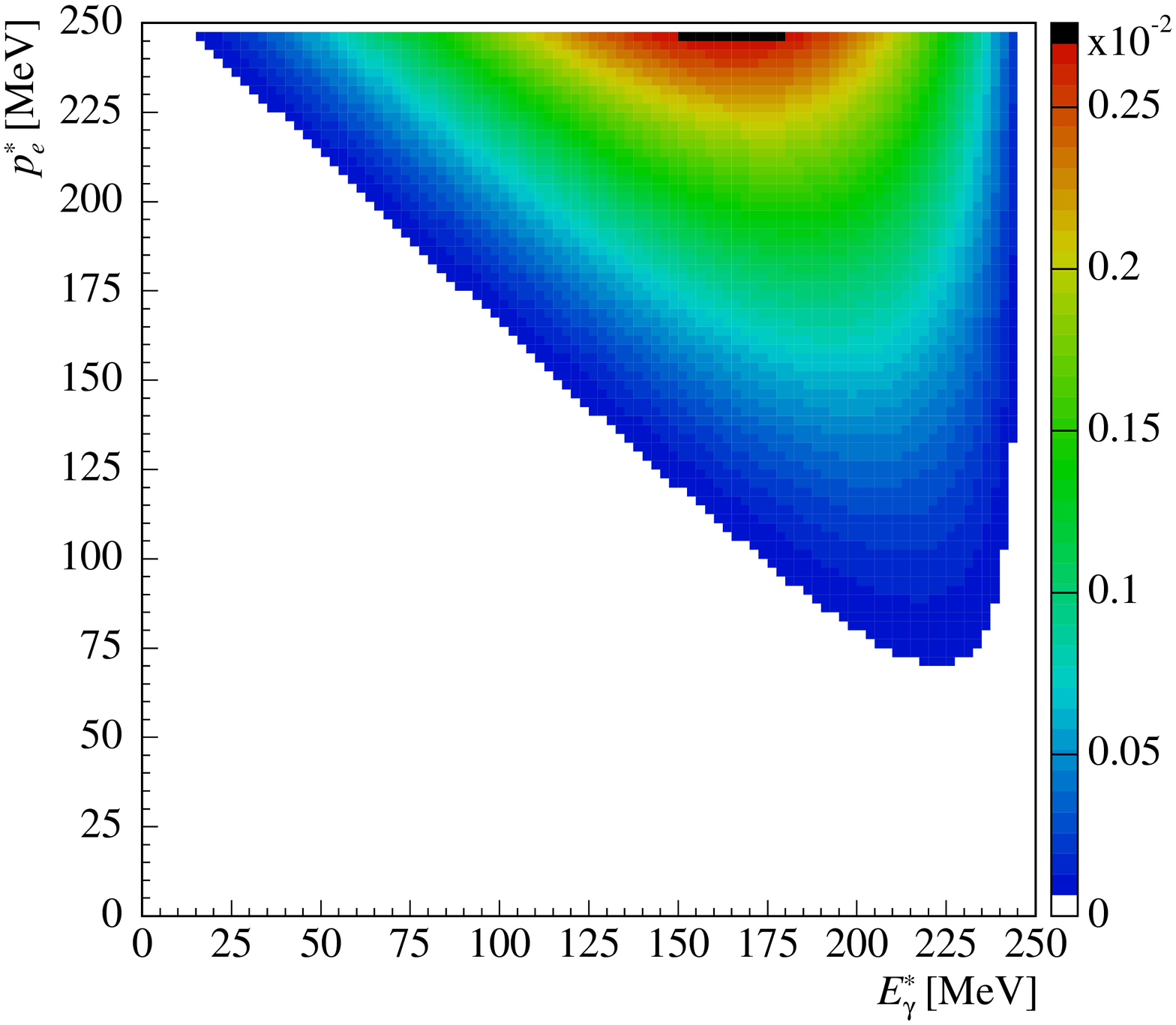}
\hspace{0.05\textwidth}
\includegraphics[width=0.33\textwidth]{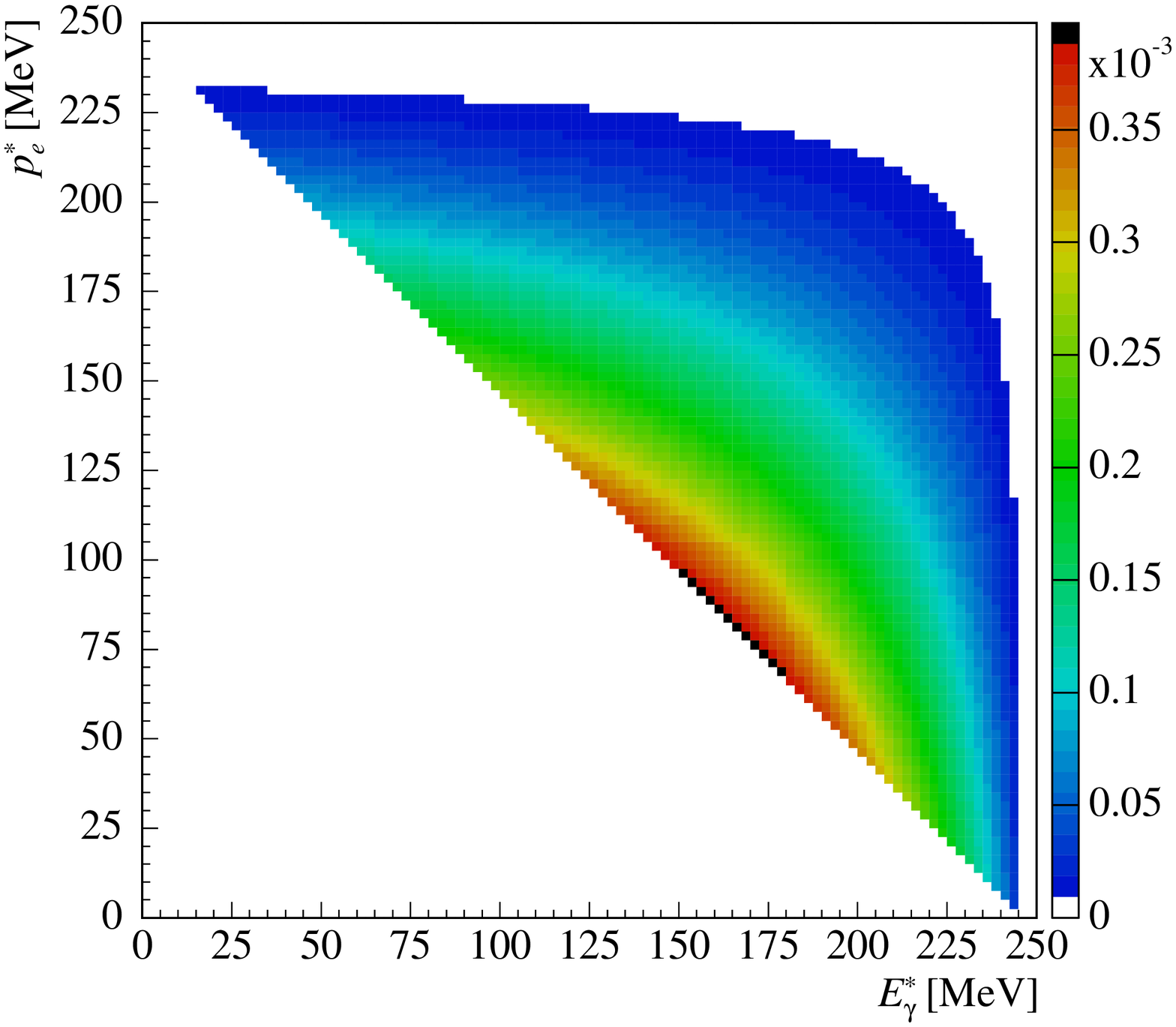}
}
\caption{Plots of expected decay intensity in the $(\Eg,\pe)$ plane 
corresponding to the contributions from the SD$+$ (left) and SD$-$ 
(right) channels, evaluated from the \order{p^4} ChPT calculation of
\Ref{B+95:l2Hand2}.}
\label{fig:SD}
\end{figure}

Decays with SD radiation can proceed through vector and axial transitions, 
with effective couplings $V$ and $A$, respectively:
\begin{equation}
\frac{d^2 \Gamma_{\rm SD}(K_{e2\gamma})}{dx\,dy} =
\frac{G_F^2\left|V_{us}\right|^2\alpha_{\rm em} M_K^5}{64\pi^2} \times 
\left[(V+A)^2f_{\rm SD+}(x,y)+(V-A)^2f_{\rm SD-}(x,y)\right].
\label{eq:SDpm}
\end{equation}
Here, $x=2\Eg/M_K$ and $y=2E_e^*/M_K$ are the normalized photon and 
electron energies in the kaon rest frame; both lie between 0 and 1.
The form factors $f_{\rm SD+}$ and $f_{\rm SD-}$ describe the 
hadronic-structure-dependent effects in the contributions to the decay
intensity from the channels with positive and negative photon polarization,
respectively. In the SD$+$ channel, the photon is preferentially emitted in 
the direction opposite to that of the $e^\pm$, and the $e^\pm$
momentum spectrum is hard,
peaking above the endpoint of the \pe\ spectrum in $K_{e3}$ decays at 
228~MeV. ($p^*_e$ is the $e^\pm$ momentum in the decay rest frame.) 
In the SD$-$ channel, the photon is preferentially emitted in 
the direction parallel to that of the $e^\pm$, and the $e^\pm$ 
has a softer momentum spectrum. This is illustrated in the Dalitz plots 
for each of the two contributions in \Fig{fig:SD}. Because the contribution
from the SD$-$ channel is completely submersed in $K_{e3(\gamma)}$ background, 
the present measurement is exclusively focused on the SD$+$ contribution,
and events with $\pe < 200$~MeV are eliminated to reduce background.

The starting sample of $K_{e2(\gamma)}$ events (inclusive with respect to
the presence of additional photons in the final state) is the same as that
used in the KLOE measurement of $R_K$ \cite{KLOE+09:Ke2,Sci09:Kaon}.
Single-prong vertices in the drift chamber
are examined if the inbound track
originates from the interaction point and has momentum compatible with
the two-body decay momentum in $\phi\to K^+ K^-$ (127~MeV), and if the
outbound track has $p_{\rm lep} > 180$ MeV.
The mass of the outbound track (the lepton candidate) is then reconstructed as 
$M^2_{\rm lep} = (E_K - |{\bf p}_{\rm miss}|)^2 - p^2_{\rm lep}$, 
with $E_K$ the kaon energy and ${\bf p_{\rm miss}}$ the missing momentum 
at the vertex. Stringent track quality cuts on both the inbound and 
outbound tracks increase the ratio of $K_{e2(\gamma)}$ events to background
events from 1/1000 to 1/20. 
For the analysis of $R_K$, the lepton-candidate track is extrapolated to the
KLOE electromagnetic calorimeter, and a neural network is used to combine
$E/p$, time-of-flight, and longitudinal energy deposition information
for the purposes of $\mu/e$ separation. 

For the analysis of $K_{e2\gamma}$, a hard cut is made on the neural-network
output. The radiative photon must be explicitly detected as a cluster not 
associated to any track, with $E_\gamma^{\rm cal} > 20$~MeV as measured in 
the calorimeter. This reduces background from activity in accidental 
coincidence with the event and from cluster fragments. Furthermore, 
the times of arrival of the lepton and photon clusters must be consistent
with each other (given the event topology) to within $2\sigma$, or about 
100~ps. This rejects $K\to\pi\pi^0$ events, in which the ``lepton'' is
actually a pion, and travels with $\beta=0.8$ instead of $\beta\approx1$,
so that its cluster arrives late.
It also rejects $K_{\mu2}$ events with accidental clusters, since the arrival
times of such clusters are distributed evenly.

\begin{figure}
\centerline{
\includegraphics[width=0.33\textwidth]{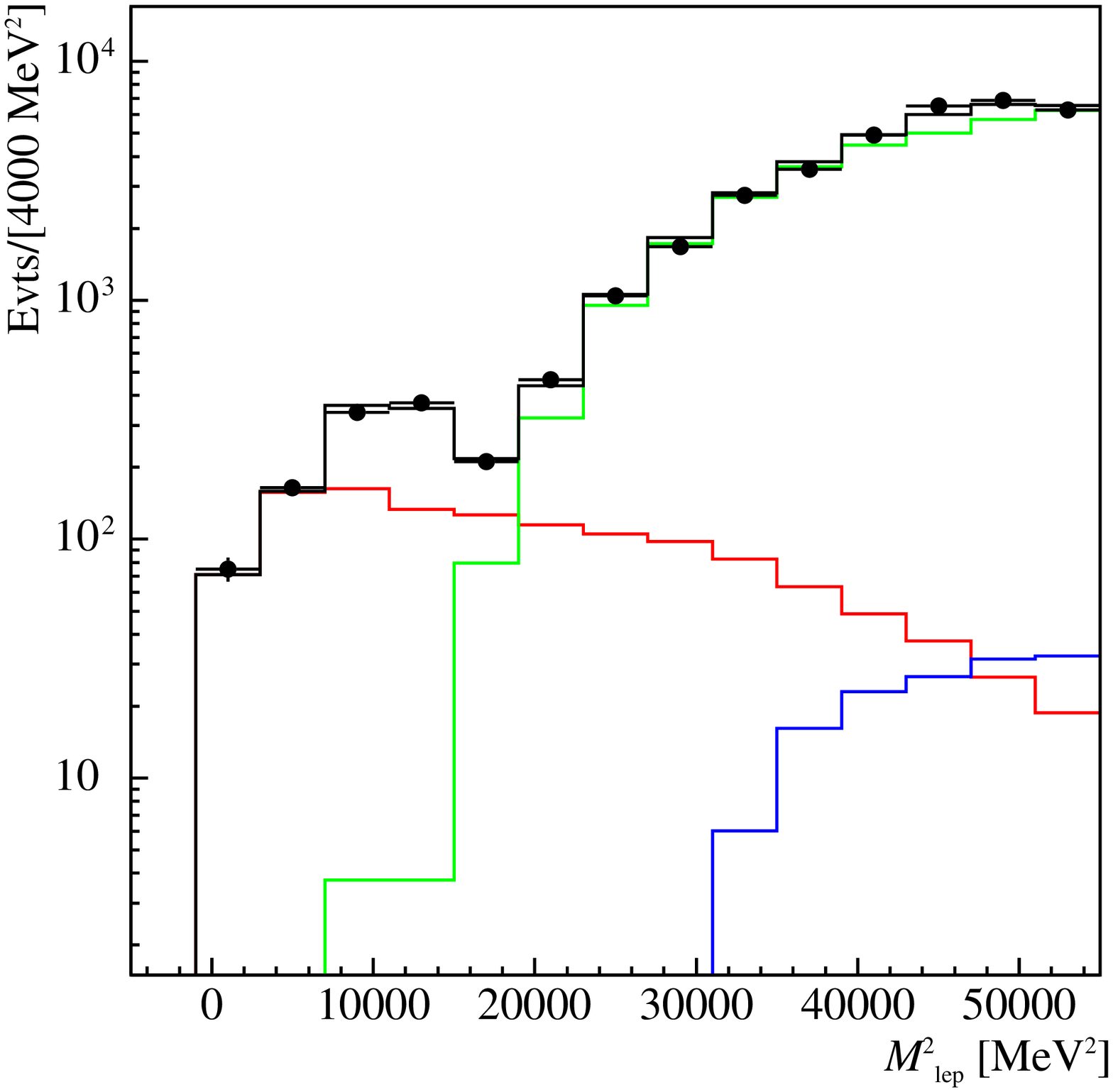}
\hspace{0.05\textwidth}
\includegraphics[width=0.33\textwidth]{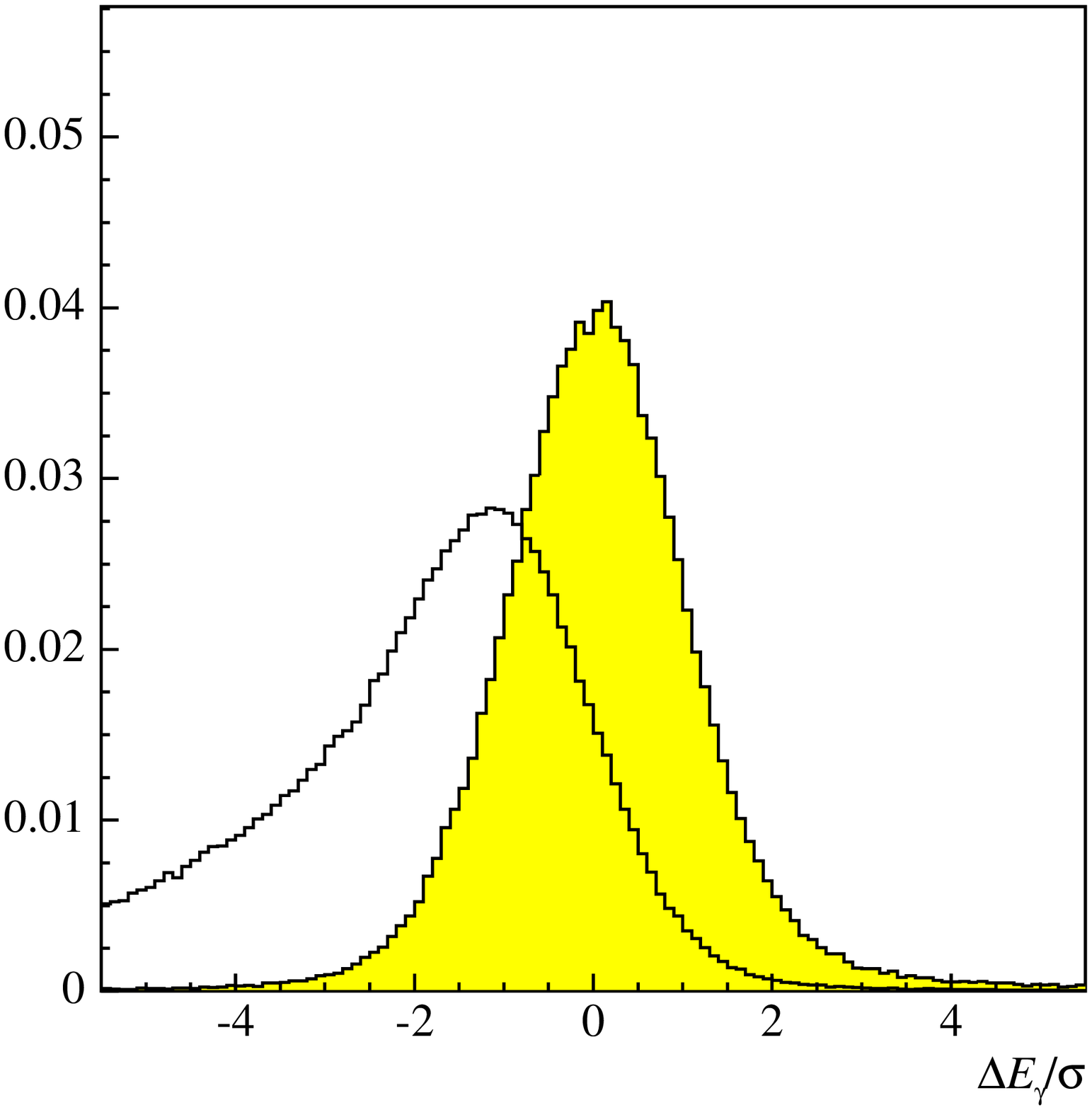}
}
\caption{Left: Distributions of $M_{\rm lep}^2$ for $K_{e2\gamma}$ candidates 
and backgrounds. The distributions for simulated $K_{e2\gamma}$ events 
with $\Eg > 10$~MeV and $\pe > 200$~MeV ($\pe < 200$~MeV) are 
shown in red (blue). The expected contribution from $K_{e3(\gamma)}$ events
is shown in green; that from all sources combined is shown in black. 
The peak at $M^2_{\rm lep} \approx 10000~{\rm MeV}^2 \approx m_\mu^2$ is from 
$K_{\mu2}$ events. The points with error bars are the data. Right: 
$\Delta E_\gamma/\sigma$ for simulated $K_{e2\gamma}$ events (yellow 
shaded histogram) and $K_{e3(\gamma)}$ events (open histogram).}
\label{fig:sig}
\end{figure}
After the above cuts, the dominant backgrounds are from $K_{\mu2(\gamma)}$
events for $M_{\rm lep}^2<20000$~MeV$^2$, and from $K_{e3(\gamma)}$ events 
for $M_{\rm lep}^2>20000$~MeV$^2$. $K_{e2\gamma}$ events with $\pe < 200$~MeV are 
submerged in background, as suggested by the left panel of \Fig{fig:sig}. 
For the purposes of this analysis, the signal sample of SD$+$ events 
is fiducially defined as the sample of identified $K_{e2\gamma}$ events
with $\Eg>10$~MeV, $\cos \qg < 0.9$ (\qg\ is the angle of the photon 
momentum with respect to that of the $e^\pm$), and 
$\pe > 200$~MeV. Simulations indicate that this selection has a 
90\% acceptance for true SD$+$ events, a 2\% acceptance for SD$-$ events,
and contains a 1\% background of IB events.

An additional variable that is useful in isolating signal events is 
$\Delta E_\gamma$, the difference between the energy of the radiative 
photon as determined by kinematic closure using the tracks in the event,
and that from the direct measurement of the energy with the calorimeter.
On average, the photon energy is determined with a resolution
of about 12 MeV in the former case and about 30 MeV in the latter.
Since the $K_{e2\gamma}$ kinematics are built into this variable, it is
particularly useful in separating $K_{e2\gamma}$ from $K_{e3(\gamma)}$ events,
as seen in the right panel of \Fig{fig:sig}.
  
The number of $K_{e2\gamma}$ events in the SD$+$ sample is determined
via log-likelihood fits to the distribution of candidate events in 
the ($M_{\rm lep}^2$, $\Delta E_\gamma$) plane in five different bins in \Eg.
The first bin in \Eg\ extends from 10 to 50~MeV; the remaining four
adjacent bins are 50 MeV wide.
The fit function is a linear combination of Monte Carlo (MC)
distributions for $K_{e2\gamma}$ events with kinematics as generated 
satisfying the SD$+$ selection (signal events), $K_{e2\gamma}$ events as
generated not satisfying the SD$+$ selection (dominantly IB events),
$K_{\mu2(\gamma)}$ events (including those with background photons in accidental
coincidence), and events from other background sources 
(mainly $K_{e3(\gamma)}$). All events in the input distributions satisfy 
the analysis cuts for the SD$+$ selection as reconstructed.
The free parameters of the fit are the weights for each 
distribution.

\begin{figure}
\centerline{
\includegraphics[width=0.3\textwidth]{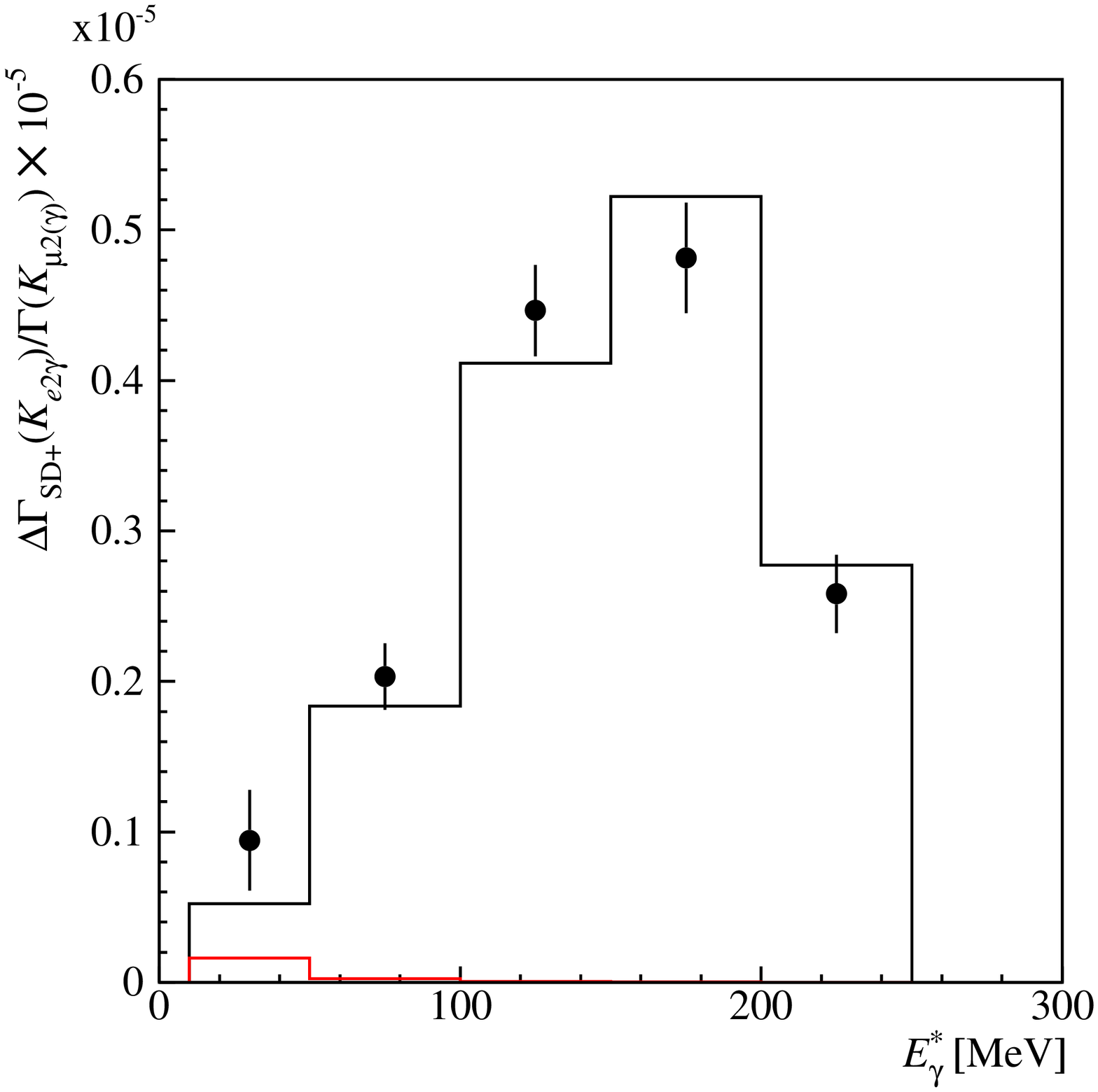}
\hfill
\includegraphics[width=0.3\textwidth]{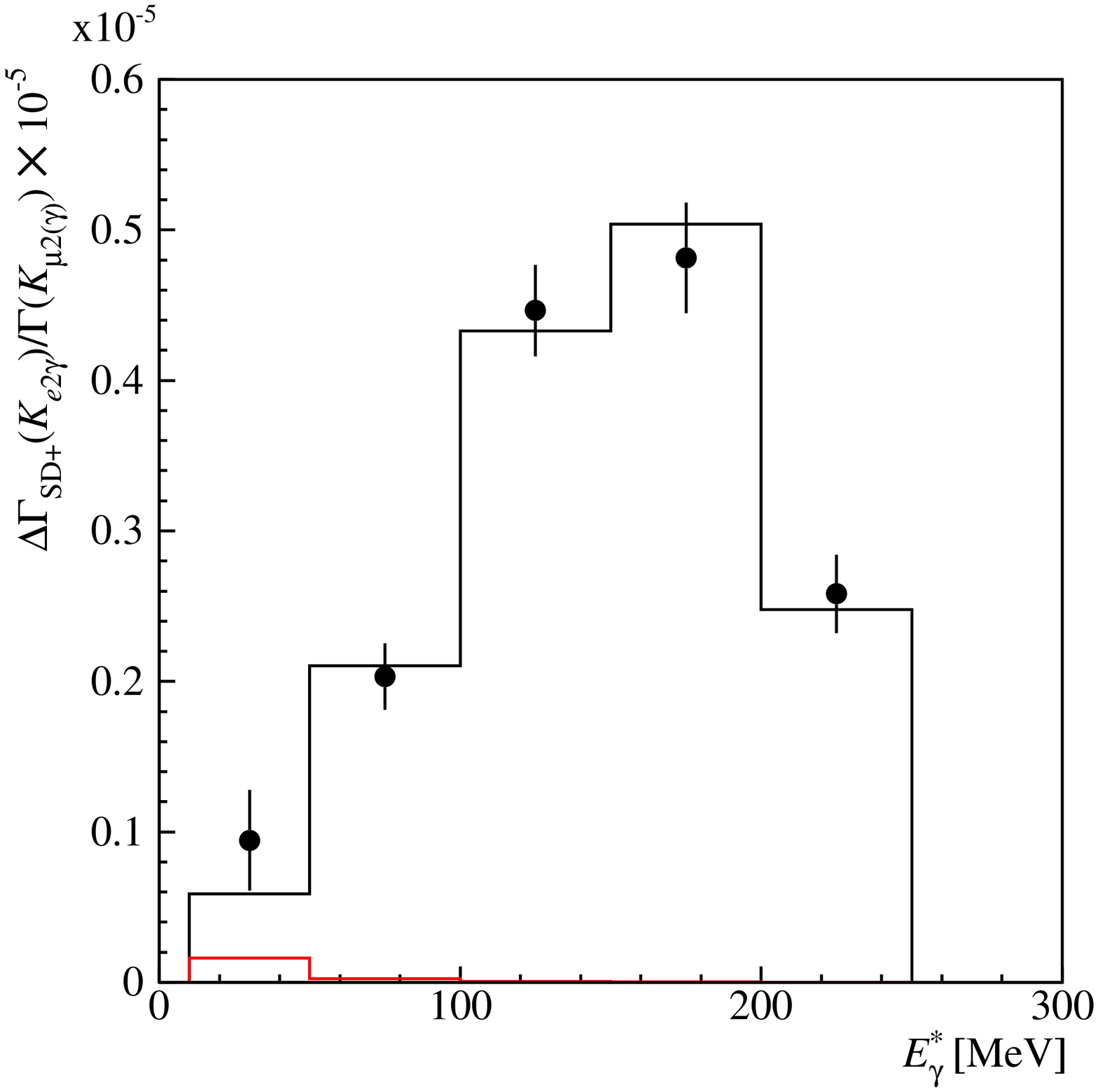}
\hfill
\includegraphics[width=0.3\textwidth]{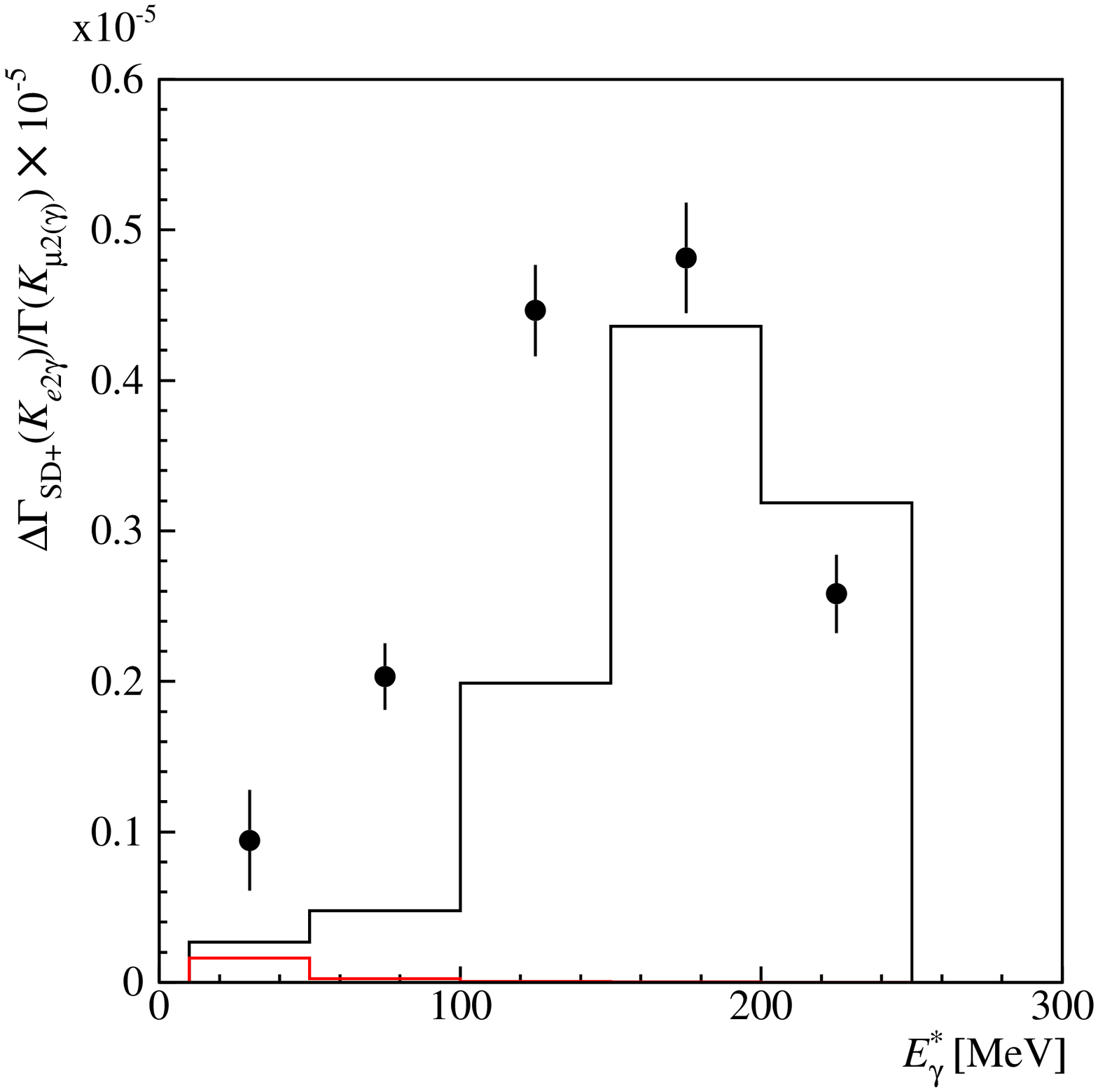}
}
\caption{$\Delta\Gamma_{\rm SD+}(K_{e2\gamma})/\Gamma(K_{\mu2(\gamma)})$
for five bins in \Eg, with model predictions.
Left: Overlay (not a fit) of \order{p^4} ChPT prediction.
Middle: Fit with \order{p^6} ChPT prediction to obtain normalization and 
slope of vector form factor.
Right: Overlay (not a fit) of LFQM prediction with parameters as in 
\Ref{CGL08:LFQM}. Red histograms show expected contribution from IB.}
\label{fig:model}
\end{figure}

Figure~\ref{fig:model} summarizes the results of this analysis.
The points in each panel show the measured values of 
$\Delta\Gamma_{\rm SD+}(K_{e2\gamma})$, 
normalized to $\Gamma(K_{\mu2(\gamma)})$,
for each of the five bins in \Eg.
Summed over all bins in \Eg, $N_{\rm SD+}(K_{e2\gamma})=1378\pm63$,
which gives $\Gamma_{\rm SD+}(K_{e2\gamma})/\Gamma(K_{\mu2(\gamma)})=
\SN{1.484(66)_{\rm st}(16)_{\rm sy}}{-5}$. This agrees with the
prediction from the KLOE MC, \SN{1.447}{-5}, which is based on the
\order{p^4} ChPT calculation of \Ref{B+95:l2Hand2},
in which $V$ and $A$ in \Eq{eq:SDpm} are the constant values
$0.095/m_K$ and $0.043/m_K$. 
In the left panel of the figure, the \order{p^4} result is overlaid with
the data; as an index of the agreement, $\chi^2/{\rm ndf} = 5.4/5$.
On the basis of this agreement, the KLOE MC is determined to be accurate in
its description of the SD radiation in $K_{e2\gamma}$ decay to within the 4.6\%
relative uncertainty of the measurement. This reduces the corresponding
contribution to the relative uncertainty on the value of $R_K$ to 0.2\% 
\cite{KLOE+09:Ke2,Sci09:Kaon}.

At \order{p^6}, the vector coupling $V$ in \Eq{eq:SDpm} depends 
linearly on the photon energy $x$, with $V = V_0[1 + \lambda(1-x)]$,
$V_0 = 0.082/m_K$ and $\lambda = 0.4$; the axial coupling
is constant, $A \approx 0.034/m_K$ \cite{CGL08:LFQM}. 
The middle panel of \Fig{fig:model} shows
the data together with the results of a fit using the \order{p^6}
prediction. Since there is no sensitivity to the SD$-$ component of 
the radiation, the quantity $(V-A)$ in \Eq{eq:SDpm} is fixed to its \order{p^6}
numerical value, and the fit parameters are $(V_0+A)$ and $\lambda$.  
The results $(V_0+A) = 0.125(7)/m_K$ and $\lambda = 0.38\pm0.21$ with 
$\rho = -0.93$ and $\chi^2/{\rm ndf} = 1.97/3$ are obtained.  
The right panel of \Fig{fig:model} shows
the data together with the results of a fit incorporating the
light front quark model (LFQM) \cite{CGL08:LFQM}, which features a complicated
dependence of $V$ and $A$ on $x$, with $V=A=0$ at $x=0$.
The data are clearly in disagreement with the prediction of this model,
as the overlay gives $\chi^2/{\rm ndf}=127/5$.

\section{$K_L\to \pi^\mp e^\pm\nu(\bar\nu)\gamma$ $(K_{e3\gamma})$}

The study of the $K_{e3\gamma}$ decay provides the opportunity to
quantitatively test predictions from ChPT. In addition, precision
measurements of the fully inclusive $K_{\ell3(\gamma)}$ decay rates 
are needed for the determination of the CKM matrix element 
$|V_{us}|$. Such measurements require an accurate understanding
of the radiation in these decays.

As in the case of $K_{e2\gamma}$ decays, the relevant kinematic variables 
for the study of $K_{e3\gamma}$ decays are \Eg\ and \qg.
The IB amplitudes diverge for $\Eg \to 0$.
The IB spectrum in \qg\ is peaked near zero as well, since
$m_e \approx 0$.
The IB and SD amplitudes interfere.
The contribution to the width from IB-SD interference is 1\% or less of the 
purely IB contribution; the purely SD contribution is negligible.

In the \order{p^6} ChPT treatment of \Ref{G+05:Ke3g}, SD radiation
is characterized by eight amplitudes, \{$V_i$, $A_i$\}, 
which in the one-loop approximation are real 
functions. These terms have similar photon energy
spectra, with maxima around $\Eg = 100$~MeV. 
This suggests the following decomposition of the photon spectrum:
\begin{equation}
  \frac{d\Gamma}{d\Eg}  =  \frac{d\Gamma_{\rm IB}}{d\Eg} + 
  \sum_{i=1}^4 \left( \langle V_i \rangle \frac{d \Gamma_{V_i}}{d\Eg} + 
  \langle A_i \rangle \frac{d \Gamma_{A_i}}{d\Eg} \right)
                           \simeq 
  \frac{d\Gamma_{\rm IB}}{d\Eg} + \mean{X} f(\Eg).
  \label{eq:x}
\end{equation}
The function $f(\Eg)$ summarizes the effect of the SD amplitude;
the parameter \mean{X} measures its strength.
The \order{p^6} ChPT estimate is $\mean{X} = -1.2 \pm 0.4$.
The low-energy constants (LECs) for the \order{p^6} terms
are unknown. An educated guess of their size leads to the assignment
of an uncertainty on \mean{X} of 30\% of the \order{p^4} 
result.

KLOE has studied the double differential rate in $K_{e3\gamma}$ decays,
$d^2\Gamma/d\Eg\,d\qg$, and measured the ratio $R$, conventionally defined as
\begin{equation}
  R \equiv \frac{\Gamma(K_{e3\gamma}; \Eg > 30~{\rm MeV}, \qg > 20^\circ)}
                {\Gamma(K_{e3(\gamma)})},
  \label{eq:ratio}
\end{equation}
The value of this ratio has been computed at 
\order{p^6} in ChPT, leading to the prediction \cite{Kub:private} 
\begin{equation}
R = \SN{(0.963 + 0.006\,\mean{X} \pm 0.010)}{-2}.
\label{eq:rx}
\end{equation}
For $\mean{X} = -1.2$, $R = \SN{(0.96\pm0.01)}{-2}$, as quoted in 
\Ref{G+05:Ke3g}.
The simultaneous measurement of $R$ and $\mean{X}$ allows a precise 
comparison with the theory, in large part avoiding complications
from the uncertainties on the LECs for \order{p^6}.

A first attempt to measure both $R$ and the magnitude of the SD 
contribution was performed by the KTeV collaboration in 2001
\cite{KTeV+01:Kl3g}. They obtained 
$R = \SN{(0.908\pm0.008^{+0.013}_{-0.012})}{-2}$;
the analysis of the SD radiation was complicated by the use of a 
cumbersome theoretical framework. The KTeV data were subsequently
reanalyzed using more restrictive cuts that provide better control over
systematic effects, but which reduce the statistics by a factor of three.
The more recent KTeV result is $R = \SN{(0.916\pm0.017)}{-2}$ 
\cite{KTeV+05:Kl3g}. No further attempt was made to isolate the SD component. 
In 2005, NA48 measured  
$R = \SN{(0.964\pm0.008^{+0.011}_{-0.009})}{-2}$ \cite{NA48+05:Ke3g}, again
without attempting to isolate the SD radiation.

The KLOE MC generates only radiation from IB, so
the \order{p^6} generator of \Ref{G+05:Ke3g} was
used for the present analysis.
The KLOE generator \cite{Gat06:rad}, makes use of
a resummation in the soft-photon limit to all orders in $\alpha$
of the \order{p^2} amplitude for single-photon emission. It is claimed
to describe the IB photon spectrum at the level of $\sim$1\%. 
While this is appropriate for inclusive decay-rate measurements 
at the 0.1\% level, the SD contribution is about 1\% of the IB contribution,
so that the accuracy level of the KLOE IB generator is of about the 
same order as the SD contribution itself.
Therefore, in this analysis, the generator of \Ref{G+05:Ke3g}
is used to obtain the photon spectrum from IB as well as from SD.

The analysis is fully described in \Ref{KLOE+08:Ke3g}. 
The criteria used to select an inclusive sample of $K_{e3\gamma}$ events 
are the same described in \Ref{KLOE+06:Ke3FF}.
Vertices in the drift chamber along an expected line of flight
reconstructed from a $K_S\to\pi^+\pi^-$ decay are identified as 
candidate $K_L$ decays.
Loose kinematic cuts are applied to remove background from 
$K_L\to\pi^+\pi^-\pi^0$ and $K_L\to\pi^+\pi^-$.
A large amount of $K_{\mu3(\gamma)}$ background is rejected using
the variable $\Delta_{\pi\mu}$, the lesser value of $|E_{\rm miss} - p_{\rm miss}|$
calculated in the $\pi\mu$ and $\mu\pi$ track mass hypotheses.
Events are retained only if $\Delta_{\pi\mu} > 10$~MeV.
The difference between the times of flight of the hypothetical
$e^\pm$ and $\pi^\mp$ tracks is used to purify the sample
to about 99.3\%.

Signal $K_{e3\gamma}$ events are selected from within the inclusive
sample. Events are first required to contain a calorimeter cluster not
associated with any track. The arrival time of this cluster must be 
within $8\sigma$ of the time expected on the basis of the position
of the decay vertex and the photon time of flight.
If there is more than one photon candidate, that with the most
compatible timing is used. Once the photon has been identified, 
the requirement $\qg>20^\circ$ is enforced. 
The photon energy in the laboratory system is calculated from
the track momenta and the photon cluster position, using the 
constraint that $m_\nu=0$. The resolution obtained is $\sim$1~MeV,
or about a factor of ten better than that of the energy measurement
from the calorimeter.
To check the agreement between MC and data on the reconstruction 
performance for quantities used for photon selection and to 
correct the MC efficiency, $K_L\to\pi^+\pi^-\pi^0$ events are used 
as a control sample.

Neural-network discriminators are used to reduce
backgrounds from $K_L\to\pi^+\pi^-\pi^0$ and $K_{\mu3(\gamma)}$.
The neural network used to identify $K_L\to\pi^+\pi^-\pi^0$ events
makes use of \Eg, \qg, the track momenta, the missing momentum, 
and $M^2_{\gamma\nu}$, the invariant mass of the photon-neutrino pair. 
The neural network used to identify $K_{\mu3(\gamma)}$ events is based on
the track momenta, the calorimeter energy measurement, and the depth of
penetration into the calorimeter of the cluster centroid.

\begin{figure}
\centerline{
\includegraphics[width=0.6\textwidth]{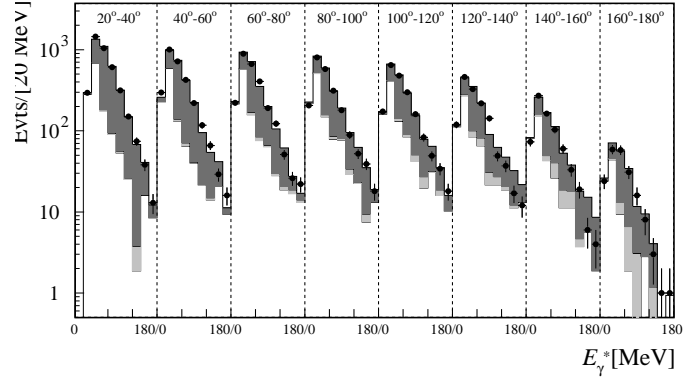}
}
\caption{
Results of fit to $(\Eg,\qg)$ distribution: dots show data,
dark gray region shows contribution from $K_{e3\gamma}$ events (IB and SD) 
satisfying kinematic cuts, white region shows contribution from $K_{e3\gamma}$
events not satisfying cuts, light gray region shows contribution from
background.
}
\label{fig:ke3fit}
\end{figure}

To count the numbers of $K_{e3\gamma}$ events from IB and SD,
a fit to the experimental distribution in $(\Eg,\qg)$ is performed 
using the sum of four independently normalized MC distributions:
the distribution for $K_{e3\gamma}$ events from IB that satisfy 
the 
cuts on \Eg\ and \qg\ as generated; 
the distribution corresponding to the function $f(\Eg)$ in the
second term of \Eq{eq:x}, representing the modification of the 
spectrum from SD events;
$K_{e3\gamma}$ events from IB that satisfy the kinematic
cuts only as reconstructed (and not as generated); and
physical background from $K_L\to\pi^+\pi^-\pi^0$ and $K_{\mu3(\gamma)}$
events.
The free parameters of the fit are the number of IB events,
the effective number of SD events (the integral of the spectral 
distortion induced by the IB-SD interference), 
and the number of $K_{e3\gamma}$ events not satisfying the kinematic cuts.
The magnitude of the background contribution is fixed using the MC.
Figure~\ref{fig:ke3fit} shows the result of the fit. 
The two-dimensional distributions are plotted on a single axis; 
the \Eg\ distributions for each of the eight slices in \qg\ 
are arrayed sequentially.
The fit gives
$\chi^2/{\rm ndf}=60/69$ ($P = 77\%$), with
$N_{\rm IB}(K_{e3\gamma}) = 9083\pm213$, 
$N_{\rm SD}(K_{e3\gamma}) = -102\pm59$, and an additional 
$6726\pm194$ $K_{e3\gamma}$ feed-down events not satisfying the
kinematic cuts. 
The negative value for the effective number of SD events 
is a result of the destructive interference between the IB and SD amplitudes.
The value $R = \SN{924(23)_{\rm st}(16)_{\rm sy}}{-5}$ is obtained.
\begin{figure}
\centerline{\includegraphics[width=0.4\textwidth]{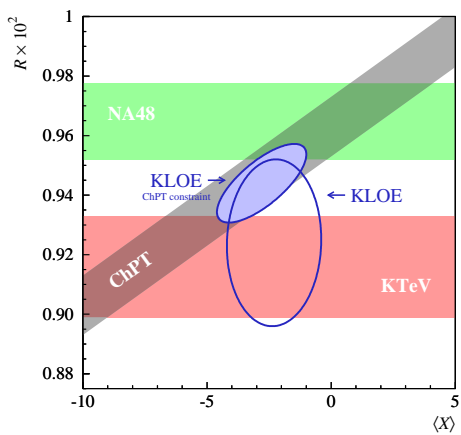}}
\caption{
KLOE $1\sigma$ contours in the $(R,\mean{X})$ plane from fit to the 
$(\Eg,\qg)$ distribution (open ellipse), and same results 
when combined with constraint from ChPT (filled ellipse). 
Results from KTeV \cite{KTeV+05:Kl3g} and NA48 \cite{NA48+05:Ke3g}
are also shown, as well as the dependence of $R$ on \mean{X} used as
the constraint.
}
\label{fig:rx}
\end{figure}
The value of the parameter \mean{X} defined in \Eq{eq:x} is derived from 
the result of the fit, taking into account the small difference in the overall 
detection efficiencies for IB and SD events.
The result is
$\mean{X} = -2.3 \pm 1.3_{\rm st} \pm 1.4_{\rm sy}.$
The $1\sigma$ contour is illustrated as the open ellipse in \Fig{fig:rx}.

The dependence of $R$ on \mean{X} of \Eq{eq:rx} is shown in 
\Fig{fig:rx} as the diagonal shaded band. This dependence can be
used to further constrain the possible values of $R$ and \mean{X} 
from this measurement, giving 
the $1\sigma$ contour illustrated as the filled ellipse 
in the figure. The constraint is applied via a fit, which gives
$R = \SN{944(14)}{-5}$ and
$\mean{X} = -2.8\pm1.8$, 
with correlation $\rho = 72\%$ and 
$\chi^2/{\rm ndf} = 0.64/1$ ($P = 42\%$).
This result represents an improved test of ChPT with respect to that 
obtained using the measurements of $R$ from 
Refs.~\cite{KTeV+05:Kl3g,NA48+05:Ke3g}.

Finally, to test the accuracy of the KLOE IB 
generator \cite{Gat06:rad}, fits to the data with no SD component
have been performed.
The results
$R = \SN{921(23)_{\rm st}}{-5}$ for the KLOE generator \cite{Gat06:rad} and
$R = \SN{925(23)_{\rm st}}{-5}$ for the \order{p^6} generator of 
\Ref{G+05:Ke3g} have been obtained.
This confirms the reliability of the KLOE generator for IB events.

\bibliographystyle{elsart-num}
\bibliography{kaon09}

\begin{thebibliography}{10}
\expandafter\ifx\csname url\endcsname\relax
  \def\url#1{\texttt{#1}}\fi
\expandafter\ifx\csname urlprefix\endcsname\relax\def\urlprefix{URL }\fi

\bibitem{CR07:Kp2}
V.~Cirigliano, I.~Rosell, Phys.\ Rev.\ Lett 99 (2007) 231801.

\bibitem{MPP06:Ke2}
A.~Masiero, P.~Paradisi, R.~Petronzio, Phys.\ Rev.\ D 74 (2006) 011701(R).

\bibitem{B+95:l2Hand2}
J.~Bijnens, et~al., in: L.~Maiani, G.~Pancheri, N.~Paver (Eds.), Second
  {DA\char8NE} Physics Handbook, Laboratori Nazionali di Frascati, 1995, p.
  315.

\bibitem{H+79:Ke2}
J.~Heintze, et~al., Nucl.\ Phys.\ B 149 (1979) 365.

\bibitem{KLOE+09:Ke2}
{KLOE~Collaboration}, F.~Ambrosino, et~al., arXiv:0907.3594 (2009).

\bibitem{Sci09:Kaon}
B.~{Sciascia for the KLOE Collaboration}, these proceedings.

\bibitem{Gou09:Kaon}
E.~{Goudzovski for the NA62 Collaboration}, these proceedings.

\bibitem{CGL08:LFQM}
C.-H. Chen, C.-Q. Geng, C.-C. Lih, Phys.\ Rev.\ D 77 (2008) 014004.

\bibitem{G+05:Ke3g}
J.~Gasser, et~al., Eur.\ Phys.\ J.\ C 40 (2005) 205--227.

\bibitem{Kub:private}
B.~Kubis, private communication.

\bibitem{KTeV+01:Kl3g}
{KTeV~Collaboration}, A.~Alavi-Harati, et~al., Phys.\ Rev.\ D 64 (2001) 112004.

\bibitem{KTeV+05:Kl3g}
{KTeV~Collaboration}, T.~Alexopoulos, et~al., Phys.\ Rev.\ D 71 (2005) 012001.

\bibitem{NA48+05:Ke3g}
{NA48~Collaboration}, A.~Lai, et~al., Phys.\ Lett.\ B 605 (2005) 247--255.

\bibitem{Gat06:rad}
C.~Gatti, Eur.\ Phys.\ J.\ C 45 (2006) 417.

\bibitem{KLOE+08:Ke3g}
{KLOE~Collaboration}, F.~Ambrosino, et~al., Eur.\ Phys.\ J.\ C 55 (2008) 539.

\bibitem{KLOE+06:Ke3FF}
{KLOE~Collaboration}, F.~Ambrosino, et~al., Phys.\ Lett.\ B 636 (2006)
  166--172.

\end{thebibliography}

\end{document}